\documentclass[10pt]{article}
\usepackage{xcolor}
\usepackage{cite}
\usepackage{amsmath,amssymb,amsfonts,amsthm}
\usepackage{algorithmic}
\usepackage{graphicx}
\usepackage{textcomp}
\usepackage{placeins}
\usepackage[left=2.5cm,right=2.5cm,top=2.5cm,bottom=3.0cm]{geometry}
\usepackage{authblk}

\usepackage{wrapfig}

\usepackage{cleveref}
\Crefformat{figure}{#2Fig.~#1#3}
\Crefmultiformat{figure}{Figs.~#2#1#3}{ and~#2#1#3}{, #2#1#3}{ and~#2#1#3}

\usepackage{dsfont}
\usepackage[T1]{fontenc}
\usepackage{graphicx}
\usepackage{amsmath,amssymb}
\usepackage{mathtools}
\usepackage{todonotes}
\usepackage{lmodern}
\usepackage{tikz}
\usetikzlibrary{spy,decorations.pathreplacing,angles,quotes,calc,arrows,positioning,}
\usepackage{multirow} 
\usepackage{pgfplots}
\pgfplotsset{compat=1.15}

\newcommand{\Id}{\mathrm{Id}}
\newcommand{\R}{\mathbb{R}}
\newcommand{\C}{\mathbb{C}}
\newcommand{\N}{\mathbb{N}}
\newcommand{\E}{\mathbb{E}}
\newcommand{\dx}{\,\mathrm{d}}
\newcommand{\SSIM}{\textrm{SSIM}}
\newcommand{\Prob}{\mathrm{p}}
\newcommand{\tr}{\operatorname{tr}}

\DeclareMathOperator*{\argmax}{argmax}
\DeclareMathOperator*{\argmin}{argmin}
\DeclareMathOperator{\prox}{prox}

\usepackage{placeins}

\definecolor{spycolor}{RGB}{150,150,200}
\tikzset{
    rectspy/.default={lens={scale=3}, size=3cm},
    rectspy on/.style={#1,},
    rectspy/.style={
        draw=spycolor,
        connect spies,
        spy scope={
        every spy on node/.style={
            draw=spycolor,
            very thick,
            rectangle, 
            rectspy on,
        },
        every spy in node/.style={
            draw=spycolor,
            very thick,
            rectangle,
        },
        #1
        },
        spy connection path={\draw[spycolor, very thick] (tikzspyonnode) -- (tikzspyinnode);}
    }
}

\tikzset{
    sepbar/.style={
        very thick,
        black!30!white,
    }
}

\newcommand{\change}[1]{#1}

\begin{document}
\title{Bayesian Uncertainty Estimation of Learned Variational MRI Reconstruction}

\author[1]{Dominik Narnhofer}
\author[2,3]{Alexander Effland}
\author[4]{Erich Kobler}
\author[5,6]{Kerstin Hammernik}
\author[7]{Florian Knoll}
\author[1]{Thomas Pock}

\affil[1]{\footnotesize Institute of Computer Graphics and Vision, Graz University of Technology, Graz, Austria}
\affil[2]{\footnotesize Silicon Austria Labs (TU Graz SAL DES Lab), Graz, Austria}
\affil[3]{\footnotesize Institute of Applied Mathematics, University of Bonn, Bonn, Germany}
\affil[4]{\footnotesize Computer Science Department, Johannes Kepler University, Linz, Austria}
\affil[5]{\footnotesize Technical University of Munich, Munich, Germany}
\affil[6]{\footnotesize Imperial College London, London, United Kingdom}
\affil[7]{\footnotesize NYU School of Medicine, New York, USA}

\date{}
\maketitle

\begin{abstract}
Recent deep learning approaches focus on improving quantitative scores of dedicated benchmarks, and therefore only reduce the observation-related (aleatoric) uncertainty.
However, the model-immanent (epistemic) uncertainty is less frequently systematically analyzed.
In this work, we introduce a Bayesian variational framework to quantify the epistemic uncertainty.
To this end, we solve the linear inverse problem of undersampled MRI reconstruction in a variational setting.
The associated energy functional is composed of a data fidelity term and 
the total deep variation (TDV) as a learned parametric regularizer.
To estimate the epistemic uncertainty we draw the parameters of the TDV regularizer from a multivariate Gaussian distribution, whose mean and covariance matrix are learned in a stochastic optimal control problem.
In several numerical experiments, we demonstrate that our approach yields competitive results for undersampled MRI reconstruction. 
Moreover, we can accurately quantify the pixelwise epistemic uncertainty, which can serve radiologists as an additional resource to visualize reconstruction reliability.
\end{abstract}


\section{Introduction}
A classical inverse problem related to magnetic resonance imaging (MRI) emerges from the undersampling of the raw data in Fourier domain (known as $k$-space) to reduce acquisition time.
When directly applying the inverse Fourier transform, the quality of the resulting image is deteriorated by undersampling artifacts since in general the sampling rate does not satisfy the Nyquist--Shannon sampling theorem.
Prominent approaches to reduce these artifacts incorporate parallel imaging~\cite{PrWe99} on the hardware side, or compressed sensing on the algorithmic side~\cite{LuDo07}.
In further algorithmic approaches, the MRI undersampling problem is cast as an ill-posed inverse problem using a hand-crafted total variation-based regularizer~\cite{KnCl12}.
In recent years, a variety of deep learning-based methods for general inverse problems have been proposed that can be adapted for undersampled MRI reconstruction, 
including deep artifact correction~\cite{YeHa18}, learned unrolled optimization~\cite{AgMa18,ScCa18,HaKl18}, or k-space interpolation learning~\cite{AkMo19}.
We refer the interested reader to~\cite{LuLu19,KnHa20} for an overview of existing methods and their applicability to MRI.

\begin{figure*}[htb]
\centering
\includegraphics[width=0.9\linewidth]{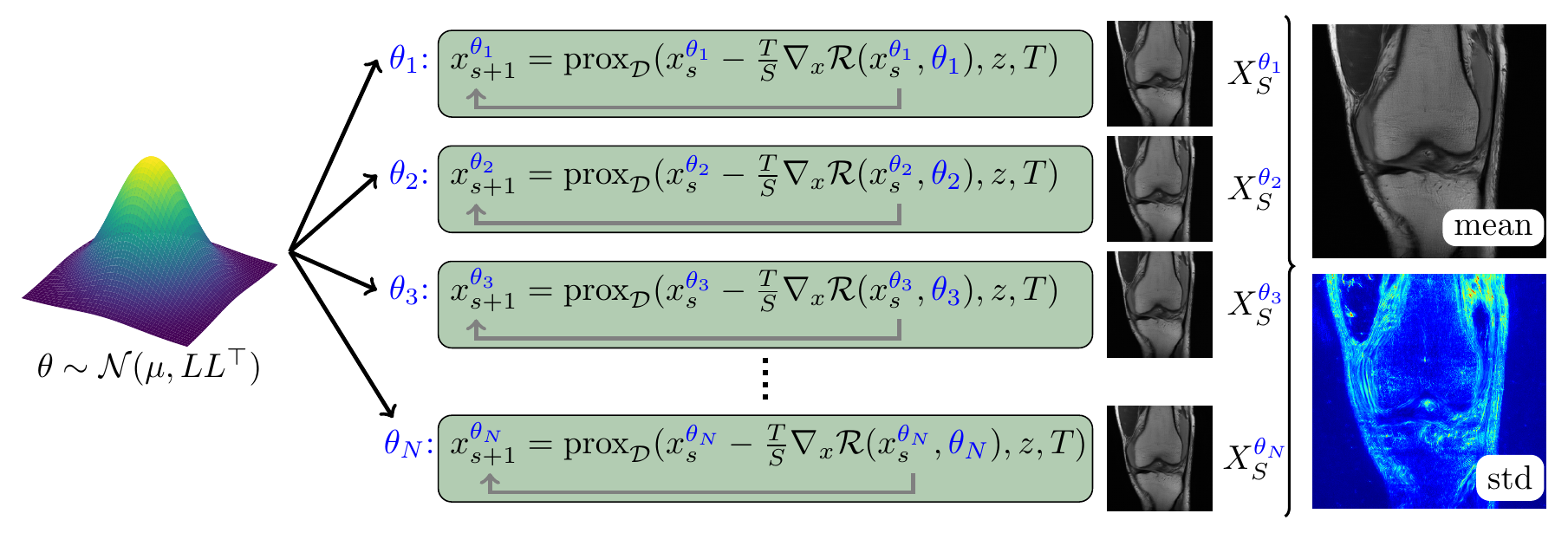}
\caption{Illustration of the stochastic MRI undersampling reconstruction model to calculate the epistemic uncertainty.
Here, $N$ instances of the model parameters $\theta$ are drawn from $\mathcal{N}(\mu,LL^\top)$,
which lead to $N$~output images $X_S^{\theta_1},\ldots,X_S^{\theta_N}$.
The associated pixelwise mean and standard deviation are depicted on the right.
}
\label{fig:uncertainty}
\end{figure*}

An established technique for solving ill-posed inverse problems are variational methods, in which the minimizer of an energy functional defines the restored output image.
A probabilistic interpretation of variational methods is motivated by Bayes' theorem, which states that the posterior distribution~$\Prob(x\vert z)$ of a reconstruction~$x$ and observed data~$z$ is proportional to $\Prob(z\vert x)\Prob(x)$.
The maximum a posteriori (MAP) estimate~\cite{Mu15} in a negative log-domain is the minimizer of the energy
\begin{equation}
\mathcal{E}(x,z)\coloneqq\mathcal{D}(x,z)+\mathcal{R}(x)
\label{eq:MAP}
\end{equation}
among all $x$, where we define the data fidelity term as $\mathcal{D}(x,z)\propto-\log(\Prob(z\vert x))$ and the regularizer as $\mathcal{R}(x)\propto-\log(\Prob(x))$.
Deep learning has been successfully integrated in this approach in a variety of papers~\cite{HaKl18,EfKo19,KoEf20}, in which the regularizer is learned from data.
However, none of these publications addresses the quantification of the uncertainty in the model itself.

In general, two sources of uncertainties exist: aleatoric and epistemic.
The former quantifies the uncertainty caused by observation-related errors, while the latter measures the inherent error of the model.
Most of the aforementioned methods generate visually impressive reconstructions and thus reduce the aleatoric uncertainty, but the problem of quantifying the epistemic uncertainty is commonly not addressed.
In practice, an accurate estimation of the epistemic uncertainty is vital for the identification of regions that cannot be reliably reconstructed such as hallucinated patterns, which could potentially result in a misdiagnosis.
This problem has been addressed in a Bayesian setting by several approaches~\cite{BlCo15,GaGh16,KeGa17,WeRo20},
but only very few are explicitly targeting uncertainty quantification for MRI. 
For instance, Schlemper et al.~\cite{ScCa18a} use Monte Carlo-dropout and a heteroscedastic loss to estimate MRI reconstruction uncertainty in U-Net/DC-CNN based models.
In contrast, Edupugantiet et al.~\cite{EdMa21} advocated a probabilistic variational autoencoder, which in combination with a Monte Carlo approach and Stein's unbiased risk estimator allows for a pixelwise uncertainty estimation.

There are two main contributions of this paper:
First, we adapt the total deep variation (TDV)~\cite{KoEf20,KoEf20a} designed as a novel framework for general linear inverse problems to undersampled MRI reconstruction, and show that we achieve competitive results on the fastMRI data set~\cite{KnZb20,KnTu20}.
In detail, we apply our method to single and multi-coil undersampled MRI reconstruction, where in the latter case no coil sensitivities are used.
Second, by roughly following the Bayes by Backprop framework~\cite{BlCo15} we pursue Bayesian inference and thus estimate the epistemic uncertainty in a pixelwise way (see \Cref{fig:uncertainty} for an illustration).
In detail, we draw the parameters of TDV from a multivariate Gaussian distribution, whose mean and covariance matrix are computed in an optimal control problem modeling training.
By iteratively drawing the parameters from this distribution we can visualize the pixelwise standard deviation of the reconstructions, which measures the epistemic uncertainty.
Ultimately, this visualization can aid clinical scientists to identify regions with potentially improper reconstructions.

\section{Methods}
In this section, we first recall the mathematical setting of undersampled MRI reconstruction.
Then, we introduce the sampled optimal control problem for deterministic and stochastic MRI reconstruction, where the latter additionally allows for an estimation of the epistemic uncertainty.

\subsection{Magnetic Resonance Imaging}
In what follows, we briefly recall the basic mathematical concepts of (undersampled) MRI.
We refer the reader to~\cite{brown2014magnetic} for further details.

Fully sampled raw data~$u\in\C^{nQ}$
are acquired in the Fourier domain commonly known as $k$-space with $Q\geq 1$~measurement coils.
Throughout this paper, the full resolution images and raw data are of size $n=\mathrm{width}\times\mathrm{height}$ and are identified with vectors in $\C^{nQ}$ (e.g.~$u\in\C^{nQ}$).
If $Q=1$, we refer to a single-coil, in all other cases to a multi-coil MRI reconstruction problem.
Here, the associated uncorrupted data in image domain representing the ground truth are given by~$y=F^{-1}u\in\C^{nQ}$,
where $F\in\C^{nQ\times nQ}$~denotes the channel-wise unitary matrix representation of the two-dimensional discrete Fourier transform and $F^{-1}$~its inverse.
In this case, the final root-sum-of-square image estimate~$Y$ for~$y$ with a resolution of $n=\text{width}\times\text{height}$ is retrieved as
\[
Y_i=\sqrt{\sum_{q=1}^Q\vert y_{i,q}\vert^2}
\]
for $i=1,\ldots,n$, where $y_{i,q}$ refers to the $i^{th}$ pixel value of the $q^{th}$~coil and $\vert\cdot\vert$ denotes the absolute value or magnitude.
Henceforth, we frequently denote the root-sum-of-square of an image by upper case letters.
Acquiring the entity of the Fourier space data (known as fully sampled MRI) results in long acquisition times and consequently a low patient throughput.
To address this issue, a subset of the $k$-space data along lines defined by a certain sampling pattern is acquired.
However, this approach violates the Nyquist--Shannon theorem, which results in clearly visible backfolding artifacts.
The aforementioned scheme is numerically realized by a downsampling operator $M_R\in\C^{(nQ/R)\times nQ}$ representing $R$-fold Cartesian undersampling ($R\in\N$), which only preserves $\frac{1}{R}$ of the lines in frequency encoding direction.
In this case, the linear forward operator is defined as $A=M_RF\in\C^{(nQ/R)\times nQ}$.
Thus, the observations resulting from the forward formulation of the inverse problem are given by 
\begin{equation}
z=Ay+\nu\in\C^{nQ/R},
\label{eq:inverseProblem}    
\end{equation}
where $\nu\in\C^{nQ/R}$ is additive noise.

\subsection{Deterministic MRI Reconstruction}
The starting point of the proposed framework is a variant of the energy formulation~\eqref{eq:MAP}.
In this paper, we use the specific data fidelity term
\[
\mathcal{D}(x,z)=\frac{1}{2}\Vert Ax-z\Vert_2^2.
\]

The data-driven regularizer~$\mathcal{R}:\R^{2nQ}\times\Theta\to\R_0^+$ depends on the learned parameters~$\theta\in\Theta\subset\R^p$, where~$\Theta$ is the space of admissible learned parameters.
Note that we use the identification $\C\cong\R^2$ to handle complex numbers.
We emphasize that in our case the regularizer is \emph{not} iteration-dependent, which implies that the learned parameters~$\theta$ are shared among all iterations leading to much fewer parameters compared to a scheme where each iteration has individual parameters.
Throughout all numerical experiments, we use the total deep variation introduced in~\cref{sub:TDV}.
Our approach is not exclusively designed for the total deep variation, which can consequently be replaced by any parametric regularizer.

In what follows, we model the training process as a sampled optimal control problem~\cite{EHa19}.
To this end, let $(y^i,z^i)_{i=1}^I\in\C^{nQ}\times\C^{nQ/R}$ be a collection of $I$~pairs of uncorrupted data~$y^i$ in image domain and associated observed $R$-fold undersampled $k$-space data~$z^i$ for $i=1,\ldots,I$, where both are related by~\eqref{eq:inverseProblem}.

Next, we approximate the MAP estimator of~$\mathcal{E}$~\eqref{eq:MAP} w.r.t.~$x$.
To this end, we use a proximal gradient scheme to increase numerical stability~\cite{ChPo16}, which is equivalent to an explicit step in the regularizer and an implicit step in the data fidelity term.
We recall that the proximal map of a function~$g$ with step size~$h>0$ is defined as
\begin{equation}
\prox_{hg}(\overline{x})=\argmin_{x}\frac{1}{2}\Vert\overline{x}-x\Vert^2_2+hg(x). 
\label{eq:DataProx}
\end{equation}
Unrolling a proximal gradient scheme on~\eqref{eq:MAP}, we obtain our model 
\begin{equation}
x_{s+1}=\prox_{\tfrac{T}{S}\mathcal{D}}(x_s-\tfrac{T}{S}\nabla_x\mathcal{R}(x_s,\theta)),
\label{eq:stateEquationDiscreteII}
\end{equation}
for $s=0,\ldots,S-1$.
Here, $S\in\N$ denotes a fixed number of iteration steps, $T>0$ is a learned scaling factor, and $\nabla_x$ denotes the gradient with respect to the $x$-component.
We define the initial state as $x_0=F^{-1}M_R^\ast z$, and the terminal state of the gradient descent~$x_S$ defines the output of our model.
The considered proximal map exhibits the closed-form expression
\[
\prox_{\tfrac{T}{S}\mathcal{D}}(\overline{x})=F^{-1}((\Id+\tfrac{T}{S}M_R^\ast M_R)^{-1}(F\overline{x}+\tfrac{T}{S}M_R^\ast z)),
\]
for which we used $A=M_RF$.
For a detailed computation we refer the reader to
\cref{sec:DataProx}.

\change{
Following~\cite{EfKo19,KoEf20,KoEf20a} we cast the training process as a discrete optimal control problem with control parameters~$T$ and $\theta$.
Optimal control theory was introduced in the machine learning community to rigorously model the training process from a mathematical perspective in~\cite{EHa19}.
Intuitively, the control parameters, which coincide with the entity of learned parameters, determine the computed output by means of the state equation.
During optimization, the control parameters are adjusted such that typically the generated output images are on average as close as possible to the respective ground truth images, where the discrepancy is quantified by the cost functional.
In our case, the state equation is given by~\eqref{eq:stateEquationDiscreteII} with initial condition $x_0=F^{-1}M_R^\ast z$.
To define the associated cost functional, we denote by $x_S(z,T,\theta)$ the terminal state of the state equation using the parameters~$T$ and $\theta$ and the data~$z$.
Furthermore, $X_S(z,T,\theta)$ defined as the root-sum-of-square of $x_S(z,T,\theta)$ coincides with the reconstructed output image.
We use the subsequent established loss functional
\begin{equation}
J(y,z,T,\theta)=\Vert X_S(z,T,\theta)-Y\Vert_1+\tau(1-\SSIM(X_S(z,T,\theta),Y))
\label{eq:costFunctional}
\end{equation}
for $\tau>0$, which balances the $\ell^1$-norm and the $\SSIM$ score.
Note that the loss functional only incorporates the difference of the magnitudes of the reconstruction~$X_S(z,T,\theta)$ and the target~$Y$.
In the cost functional given by
\begin{equation}
\inf_{T\in\R_0^+,\theta\in\Theta}
\tfrac{1}{I}\sum_{i=1}^IJ(y^i,z^i,T,\theta)
\label{eq:discreteOptimalControl}
\end{equation}
the discrepancy of the reconstructions and the targets among the entire data set is minimized.
For further details we refer the reader to the literature mentioned above.
}

\subsection{Bayesian MRI Reconstruction}
Inspired by~\cite{BlCo15}, we estimate the epistemic uncertainty of the previous deterministic model by sampling the weights from a learned probability distribution.
Here, we advocate the Gaussian distribution as a probability distribution for the parameters, which is justified by the central limit theorem and has been discussed in several prior publications~\cite{LeBa18,Fort21,GaRa18}.
For instance, according to~\cite{Wi96} a neural network with only a single layer and a parameter prior with bounded variance converges in the limit of the kernel size to a Gaussian process.
For further examples of central limit type convergence estimates for neural networks we refer the reader to the aforementioned literature and the references therein.
The second major advantage of this choice is the availability of a closed-form expression of the Kullback--Leibler divergence for Gaussian processes, which is crucial for the efficient proximal optimization scheme introduced below.

We draw the weights~$\theta$ of the regularizer from the multivariate Gaussian distribution $\mathcal{N}(\mu,\Sigma)$ with a learned mean~$\mu\in\Theta\subset\R^p$ and covariance matrix~$\Sigma\in\R^{p\times p}$.
To decrease the amount of learnable parameter, we reparametrize $\Sigma=LL^\top\in\R^{p\times p}$, where $L\in\R^{p\times p}$ is a learned lower triangular matrix with non-vanishing diagonal entries.
In particular, $\Sigma$ is always positive definite and symmetric.
For simplicity, we assume that~$\Sigma$ admits a block diagonal structure, in which the diagonal can be decomposed into blocks of size~$o^2\times o^2$.
Here, each block describes the covariance matrix of a single kernel of size~$o\times o$ of a CNN representing the regularizer.
In particular, there is no correlation among the kernel weights of different kernels, i.e. $o=3$ throughout this work.

Realizations of parameters $\theta$ can simply be computed by using the reparametrization $\theta = \mu + Lz$ for $z~\sim~\mathcal{N}(0,\Id)$.

As a straightforward approach to model uncertainty, one could minimize \eqref{eq:discreteOptimalControl} w.r.t.~$\mu$ and~$L$.
However, in this case a deterministic minimizer with $\theta=\mu$ and $\Sigma$ being the null matrix is retrieved.
Thus, to enforce a certain level of uncertainty we include the Kullback--Leibler divergence~$\mathrm{KL}$ in the loss functional~\cite{Ma03}.
We recall that $\mathrm{KL}$ for two multivariate probability distributions~$p_1$ and~$p_2$ with density functions~$f_1$ and~$f_2$ on a domain~$\Omega$ reads as
\[
\mathrm{KL}(p_1\Vert p_2)
=\int_\Omega f_1(x)\log\left(\frac{f_1(x)}{f_2(x)}\right)\dx x.
\]
In particular, $\mathrm{KL}$ is non-negative and in general non-symmetric, and can be regarded as a discrepancy measure of two probability distributions.
In the special case of multivariate Gaussian probability distributions $p_1=\mathcal{N}(\mu_1,\Sigma_1)$ and $p_2=\mathcal{N}(\mu_2,\Sigma_2)$, the Kullback--Leibler divergence admits the closed-form expression
\begin{equation}
\mathrm{KL}(p_1\Vert p_2)
=\tfrac{1}{2}\left(\log\tfrac{\vert\Sigma_2\vert}{\vert\Sigma_1\vert}+\tr(\Sigma_2^{-1}\Sigma_1)+(\mu_2-\mu_1)^\top\Sigma_2^{-1}(\mu_2-\mu_1)-d\right).
\label{eq:multivariateKLD}
\end{equation}

In this paper, we use the particular choice
\[
p_1=\mathcal{N}(\mu,\Sigma),\quad p_2=\mathcal{N}(\mu,\alpha^{-1}\Id), 
\]
where~$\mu$ and~$\Sigma=LL^\top$ are computed during optimization and 
$\alpha>0$ is an a priori given constant.
This choice is motivated by the fact that the mean of the sampled weights should be determined in the optimal control problem while the constant~$\alpha$ is essential to control the level of uncertainty in the model.

Here, smaller values of $\alpha$ enforce higher levels of uncertainty, and in the limit case $\alpha\to\infty$ the deterministic model is retrieved.

Neglecting constants and scaling the Kullback--Leibler divergence with~$\beta\geq0$ leads to the subsequent stochastic sampled optimal control problem
\begin{align}
\inf\Big(&
\E_{\theta\sim\mathcal{N}(\mu,LL^\top)}\Big[\tfrac{1}{I}\sum_{i=1}^IJ(y^i,z^i,T,\theta)\Big]\notag+\beta(\alpha\tr(LL^\top)-\log(\det(LL^\top))):\notag\\
&T\in\R_0^+,\mu\in\Theta,L\in\R^{p\times p}\text{ with }\det(L)\neq0
\label{eq:stochasticOptimization}
\Big).
\end{align}
Note that \eqref{eq:stochasticOptimization} coincides with the deterministic model if $\beta=0$.
Indeed, in this case the MAP estimate is retrieved which minimizes the functional~$J$ since no uncertainty is promoted.
In summary, $\alpha$ controls the covariance matrix of the Gaussian distribution to which the learned $\Sigma$ should be close.
The parameter~$\beta$ can be regarded as the strength of the penalization to enforce this constraint, and thus also controls the dynamics during optimization.

Next, we are concerned with the minimization of~\eqref{eq:stochasticOptimization}.
First, we observe that~\eqref{eq:stochasticOptimization} is actually composed of the non-convex loss function
\[
\mathcal{J}(T,\mu,L)\coloneqq\E_{\theta\sim\mathcal{N}(\mu,LL^\top)}\left[\tfrac{1}{I}\sum_{i=1}^IJ(y^i,z^i,T,\theta)\right]
\]
as well as the convex regularization term
\[
f(L)\coloneqq\beta(\alpha\tr(LL^\top)-\log(\det(LL^\top))).
\]
For minimizing the composite loss function~\eqref{eq:stochasticOptimization}, we again use a proximal gradient descent-based scheme.
A proper optimization crucially relies on different step sizes $h_{l_u}^k$ for each diagonal $o^2\times o^2$ block $l_u$ of~$L$ due to the different magnitudes in the correlation matrix.
Here, we used that the block diagonal structure of~$\Sigma$ translates to the corresponding structure in~$L$, which admits a decomposition into~$o^2\times o^2$ blocks.
Hence, our update scheme is given by
\[
\begin{pmatrix}
T^{k+1}\\
\mu^{k+1}\\
l_u^{k+1}
\end{pmatrix}
=
\begin{pmatrix}
T^k-h_T^k\nabla_T\mathcal{J}(T^k,\mu^k,L^k)\\
\mu^k-h_\mu^k\nabla_\mu\mathcal{J}(T^k,\mu^k,L^k)\\
\prox_{h_{l_u}^kf}(l_u^k-h_{l_u}^k\nabla_L\mathcal{J}(T^k,\mu^k,l_u^k))
\end{pmatrix}
\]
for all~$u$, where the iteration-dependent step sizes $h_T^k,h_\mu^k,h_{l_u}^k>0$ are adjusted by the ADAM optimizer~\cite{KiBa15}.
On each block~$l_u$, the proximal map is defined as
\begin{equation}
\prox_{h f}(\overline{l}_u)=\argmin_{l_u}\tfrac{1}{2h}\Vert l_u-\overline{l}_u\Vert_2^2+f(l_u),
\label{eq:proxDefinition}    
\end{equation}
where the minimum is taken among all non-singular lower triangular matrices of size~$o^2\times o^2$.
Given a regular lower triangular block matrix~$\overline{l}\in\R^{o^2\times o^2}$, the proximal map of~$f$
admits the closed-form expression
\[
\prox_{h f}(\overline{l})_{ab}=
\begin{cases}
\displaystyle\frac{\overline{l}_{aa}+\sqrt{\overline{l}_{aa}^2+8\beta h(1+2\alpha\beta h)}}{2(1+2\alpha\beta h)}
,& a=b,\\[1em]
(1+2\alpha\beta h)^{-1}\overline{l}_{ab},& a\neq b.
\end{cases}
\]
A detailed computation of this proximal map can be found in \cref{sec:KLDProx}.

Finally, we stress that $\beta$ determines the level of entropy inherent in the model.
We define that the mean entropy $\overline{H}$ as a measure of uncertainty in the model~\cite{Sh48} for the convolutional kernels~$K_1$ and~$K_2$ of all residual blocks as
\[
\overline{H}(\Sigma)=\frac{1}{2N_K}\sum_{i=1}^{N_K}\ln(2\pi\det(\Sigma_i)),
\]
where $N_K$ is the total number of stochastic convolutional kernels in the network and $\Sigma_i$ is the collection of associated covariance matrices of each kernel.

\subsection{Total Deep Variation}\label{sub:TDV}
The data-driven TDV regularizer~$\mathcal{R}(x,\theta)$, depending on the learned parameters~$\theta\in\Theta\subset\R^p$, was originally proposed in~\cite{KoEf20,KoEf20a}.
In detail, $\mathcal{R}:\R^{2nQ}\times\Theta\to\R_0^+$ is computed by summing the pixelwise regularization energy~$r:\R^{2nQ}\times\Theta\to\R^n$, i.e.~$\mathcal{R}(x,\theta)=\sum_{l=1}^nr(x,\theta)_l$,
which is defined as $r(x,\theta)=w\psi(K_0x)$.
\begin{wrapfigure}{r}{0.4\textwidth}
\includegraphics[width=\linewidth]{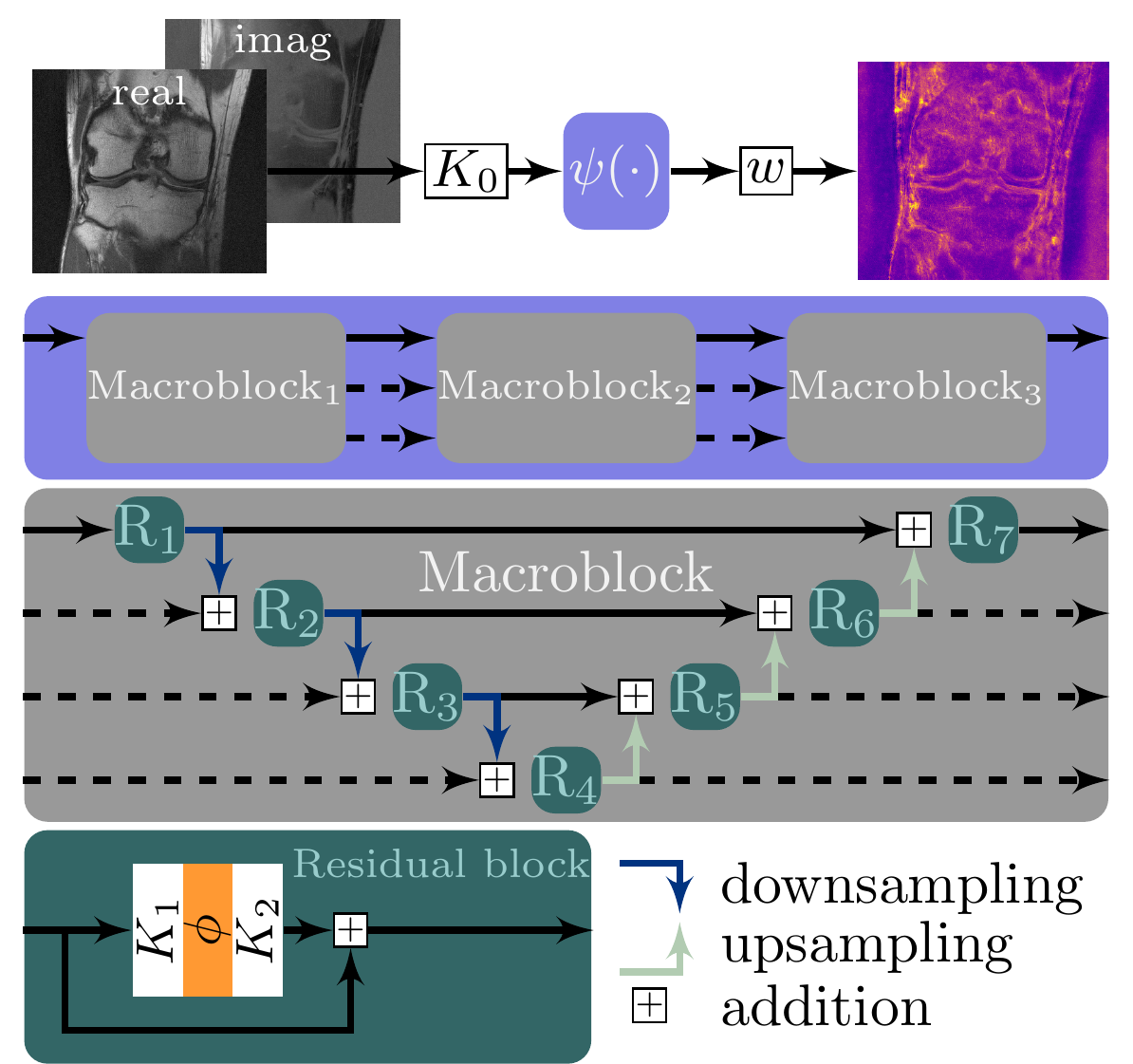}
\caption{The building blocks of the total deep variation with $3$~macroblocks (Figure adapted from~\cite[Figure~1]{KoEf20}).
Complex data are transformed to a pixelwise energy as seen in the top right corner.}
\label{fig:TDV}
\end{wrapfigure}
Note that we use the identification $\C\cong\R^2$ to handle complex numbers.
The building blocks of~$r$ are as follows:
\begin{itemize}
\item
$K_0\in\R^{nm\times 2nQ}$ is the matrix representation of a $3\times 3$ convolution kernel with $m$~feature channels and zero-mean constraint, which enforces an invariance with respect to global shifts,
\item
$\psi:\R^{nm}\to\R^{nm}$ is a convolutional neural network (CNN) described below,
\item
$w\in\R^{n\times nm}$ is the matrix representation of a learned $1\times 1$ convolution layer.
\end{itemize}
Note that $\theta$ encodes~$K_0$, and all convolutional weights in~$\psi$ and~$w$.
The CNN~$\psi$ is composed of $3$~macroblocks connected by skip connections (\Cref{fig:TDV}, second row),
where each macroblock consists of $7$~residual blocks (\Cref{fig:TDV}, third row).
We remark that the core architecture is inspired by a U-Net~\cite{RoFi15} with additional residual connections and a more sophisticated structure.
Each residual block has two bias-free $3\times 3$~convolution layers~$K_1,K_2\in\R^{nm\times nm}$ and a smooth log-student-t activation function~$\phi(x)=\frac{1}{2}\log(1+x^2)$ as depicted in~\Cref{fig:TDV} (last row).
This choice of the activation function is motivated by the pioneering work of Mumford and coworkers~\cite{HuMu99}.
The down-/upsampling are realized by learned $3\times 3$ convolution layers and transposed convolutions, respectively, both with stride~$2$.
For the stochastic setting, we henceforth assume that $T$ is fixed and that $K_0$, the down- and upsampling operators, and $w$ are always deterministic.

\subsection{Numerical Optimization and Training Data}
To optimize the optimal control problems in the deterministic~\eqref{eq:discreteOptimalControl} and stochastic~\eqref{eq:stochasticOptimization} regime, we use the ADAM algorithm~\cite{KiBa15} with a batch size of~$8$, momentum variables $\beta_1=0.5$ and $\beta_2=0.9$, where the first and second moment estimates are reinitialized after 
$50\,000$~parameter updates.
The initial learning rate is $10^{-4}$, which is halved each $50\,000$~iterations, and the total number of iterations is $120\,000$ for $Q=1$ and $200\,000$ for $Q\geq 2$.
The memory consumption is reduced by randomly extracting patches of size $96\times 368$ in frequency encoding direction as advocated by~\cite{ScCa18}.
To further stabilize the algorithm and increase training performance, we start with $S=2$ iterations, which is successively incremented by~$1$ after $7\,500$~iterations.
For the same reason, we retrain our model for $R=8$ starting from the terminal parameters for $R=4$ using $100\,000$ iterations for $Q=1$ and $130\,000$ for $Q\geq 2$.

In all experiments, we train our model with the data and the random downsampling operators of the single and multi-coil knee data of the fastMRI data set~\cite{KnZb20},
for which two different acquisition protocols were used: PD (coronal proton density scans) and PD-FS (coronal proton density scans with fat saturation).

Moreover, the undersampling pattern defining~$M_R$ is created on-the-fly by uniformly sampling lines outside the fixed auto calibration area such that only $\frac{1}{R}$ of the lines are preserved.
The associated ground truth images are computed using the emulated single-coil methodology~\cite{TyZb18} in the single-coil case and the root-sum-of-squares reconstructions in the multi-coil setting, which is consistent with the methodology in~\cite{KnZb20}.
We emphasize that no separate training for different modalities including the acquisition protocol, the field strength and the manufacturer is conducted.

Training and inference were performed on a 20~core 2.4GHz Intel Xeon machine equipped with a NVIDIA~Titan~V~GPU.
The entire training of a single example took roughly 10~days/14~days in the deterministic/stochastic case requiring 12GB of GPU memory.
In both cases, inference takes 3~seconds per sample and requires 3.2GB of GPU memory.

\subsection{Stochastic Reconstruction}
\change{
Next, we discuss how to retrieve estimates of the undersampled MRI reconstruction in the stochastic setting.
First, we integrate the learned distribution $p(\theta)$ of the parameters in the Bayesian formula by noting that $\Prob(z\vert x,\theta)=\Prob(z\vert x)$ as follows:
\[
\Prob(x,\theta\vert z)= \frac{\Prob(z\vert x) \Prob(x\vert \theta)\Prob(\theta)}{\Prob(z)}.
\]
%
By marginalizing over $\theta$ we obtain
\[
\Prob(x\vert z)=
\int_\Theta
\frac{\Prob(z\vert x)\Prob(x\vert\theta)\Prob(\theta)}{\Prob(z)}
\dx\theta.
\]
In our case, a closed-form solution of the integral is not available due to the non-linearity in the regularizer. 
However, an approximation of the integral can be obtained by Monte--Carlo sampling
\begin{equation}
\Prob(x\vert z)\approx\frac{1}{N}\sum_{i=1}^{N} \frac{\Prob(z\vert x) \Prob(x\vert \theta_i)}{\Prob(z)},
\label{eq:MonteCarlo}
\end{equation}
where $\theta_i$ is randomly drawn from the probability distribution~$\Prob(\theta)$.
We refer the reader to~\cite{Ca98,BaZh20} for the consistency of this approximation as well as the corresponding convergence rates.
}
Furthermore, for each instance $\theta_i$ we denote by $x_s(z,T,\theta^i)$ the approximate optimal solution of the variational problem
\begin{equation}
\argmax_{x\in\C^{nQ}}\Prob(z\vert x)\Prob(x\vert \theta_i)
\label{eq:XThetaI}
\end{equation}
in this setting. 
\change{Note that $p(z)$ appearing in~\eqref{eq:MonteCarlo} does not affect the maximizer, that is why we omit this term in~\eqref{eq:XThetaI}.}

\change{
To retrieve estimates of the undersampled MRI reconstruction in the stochastic setting, we draw $N\in\N$~instances $\underline{\theta}_N=(\theta_1,\ldots,\theta_N)\in\Theta^N$ from $\mathcal{N}(\mu,LL^\top)$, where $\mu$ and $L$ are determined by~\eqref{eq:stochasticOptimization}.
In a negative logarithmic domain, the maximization problem~\eqref{eq:XThetaI} is equivalent to
\[
\argmin_{x\in\C^{nQ}}\mathcal{D}(x,z)+\mathcal{R}(x,\theta_i),
\]
where we have identified the first factor with the data fidelity term and the second factor with the regularizer as above.
As before, the approximate minimizer is denoted by $x_S(z,T,\theta^i)$ and computed as in \eqref{eq:stateEquationDiscreteII}.
Then, the average~$\overline{x}^N$ and the corresponding standard deviation~$\sigma^N$ of $N$ independent realizations are defined as
\begin{align*}
\overline{x}_S^N(z,T,\underline{\theta}_N)&=\tfrac{1}{N}\sum_{i=1}^N x_S(z,T,\theta^i),\\
(\sigma_S^N(z,T,\underline{\theta}_N))_j^2&=\tfrac{1}{N}\sum_{i=1}^N((x_S(z,T,\theta^i)-\overline{x}_S^N(z,T,\underline{\theta}_N))_j)^2
\end{align*}
for each pixel $j=1,\ldots,n$.
In particular, $\overline{x}_S^N(z,T,\underline{\theta}_N)$ refers to the averaged output image.
This approach is a special form of \emph{posterior sampling} and summarized in \Cref{fig:uncertainty}.
Finally, the root-sum-of-square reconstruction~$\overline{X}_S^N(z,T,\underline{\theta}_N)$ of $\overline{x}_S^N(z,T,\underline{\theta}_N)$ and the corresponding standard deviation are given by
\begin{align}
\overline{X}_S^N(z,T,\underline{\theta}_N)&=\tfrac{1}{N}\sum_{i=1}^N X_S(z,T,\theta^i),\label{eq:meanDefintion}\\
(\widehat{\sigma}_S^N(z,T,\underline{\theta}_N))_j^2&=\tfrac{1}{N}\sum_{i=1}^N((X_S(z,T,\theta^i)-\overline{X}_S^N(z,T,\underline{\theta}_N))_j)^2,\notag
\end{align}
respectively.
}

\section{Numerical Results}
In this section, we present numerical results for single and multi-coil undersampled MRI reconstruction in the deterministic and stochastic setting.
In all experiments, we set the initial lower triangular matrix~$L_0 = \sqrt{10^{-3}}\Id$, $\alpha=10$, $S=15$ and~$N=32$, in all multi-coil results we have $Q=15$ coils.

\begin{table*}[htb]
\caption{Quantitative results for various single and multi-coil MRI reconstruction methods for $R\in\{4,8\}$.}
\centering
\resizebox{\linewidth}{!}{%
\begin{tabular}{l|l|ccc|ccc|c}
\multicolumn{2}{c}{} & \multicolumn{3}{c}{\textbf{$R=4$}} & \multicolumn{3}{c}{\textbf{$R=8$}} & \\
\cline{4-4}  \cline{7-7}
\noalign{\smallskip}
\textbf{Acquisition} & \textbf{Method} & \textbf{PSNR~$\uparrow$}  & \textbf{NMSE~$\downarrow$} & \textbf{SSIM~$\uparrow$} & \textbf{PSNR~$\uparrow$}  & \textbf{NMSE~$\downarrow$} & \textbf{SSIM~$\uparrow$} & \textbf{Parameters ($\times 10^6$)} \\  \hline\hline
\multirow{5}{*}{single-coil}  &     zero filling                &    30.5       &    0.0438         &     0.687         &           26.6     &     0.0839         &     0.543   & $-$            \\
                              &     U-Net~\cite{KnZb20}         &    32.2       &    0.032          &     0.754         &           29.5     &     0.048          &     0.651   & $214.16$       \\
                               &     $\Sigma$-Net~\cite{HaSc19} &    33.5 	    &    0.0279         &    0.777          &           n.a.     &     n.a.           &     n.a.    & 140.92       \\
                              &     iRim~\cite{PuKa20}          &    33.7 		&    0.0271         &     0.781         &           30.6     &     0.0419         &     0.687   & $275.25$       \\
                              &     TDV (deterministic)         &    33.8 	    &    0.0257         &     0.768         &           30.5     &     0.0407         &     0.665   & $2.21$         \\
                              &     TDV (stochastic)            &    33.5 		&    0.0269         &     0.762         &           30.4     &     0.0412         &     0.662   & $9.95$         \\\hline
\multirow{5}{*}{multi-coil}   &     zero filling                &    32.0	    &    0.0255         &     0.848         &           28.4     &     0.0549         &     0.778   & $-$            \\
                              &     U-Net~\cite{KnZb20}         &    35.9 	    &    0.0106         &     0.904         &           33.6     &     0.0171         &     0.858   & $214.16$       \\
                              &     $\Sigma$-Net~\cite{HaSc19}  &    39.8 	    &    0.0051         &     0.928         &           36.7     &     0.0091         &     0.888   & 675.97       \\
                              &     iRim~\cite{PuKa20}          &    39.6	    &    0.0051         &     0.928         &           36.7     &     0.0091         &     0.888   & $329.67$       \\
                              &     E2EVN~\cite{SrZb20}         &    39.9		&    0.0049         &     0.930         &           36.9     &     0.0089         &     0.890   & $30.0$    \\
                              &     TDV (deterministic)         &    39.3		&    0.0054         &     0.923         &           35.9     &     0.0108         &     0.876   & $2.22$         \\
                              &     TDV (stochastic)            &    38.9		&    0.0058         &     0.919         &           35.2     &     0.0123         &     0.867   & $9.97$         \\

\end{tabular}
}
\label{tab:PSNR}
\end{table*}

\subsection{MRI Reconstruction}
\Cref{tab:PSNR} lists quantitative results for $R\in\{4,8\}$ of the initial zero filling, two state-of-the-art methods (U-Net~\cite{KnZb20} and iRim~\cite{PuKa20}, values taken from the public leaderboard of the fastMRI challenge (for further details see \underline{https://fastmri.org/leaderboards}) for both the deterministic and stochastic version of our approach.
We stress that we jointly train our model for all contrasts without any further adaptions.
In particular, we did not incorporate any metadata in the training process such as contrast levels, manufacturer or field strength. 
Moreover, our model exhibits an impressively low number of parameters compared to the competing methods, which have up to $300$~times more parameters.
Note that although the number of trainable parameters in the stochastic TDV is larger compared to the deterministic version, the number of sampled parameters $\theta$ used for reconstruction is identical to the deterministic case.
All MRI reconstructions are rescaled to the interval~$[0,1]$ to allow for an easier comparison.

\begin{figure*}[htb]
\includegraphics[width=1.0\linewidth]{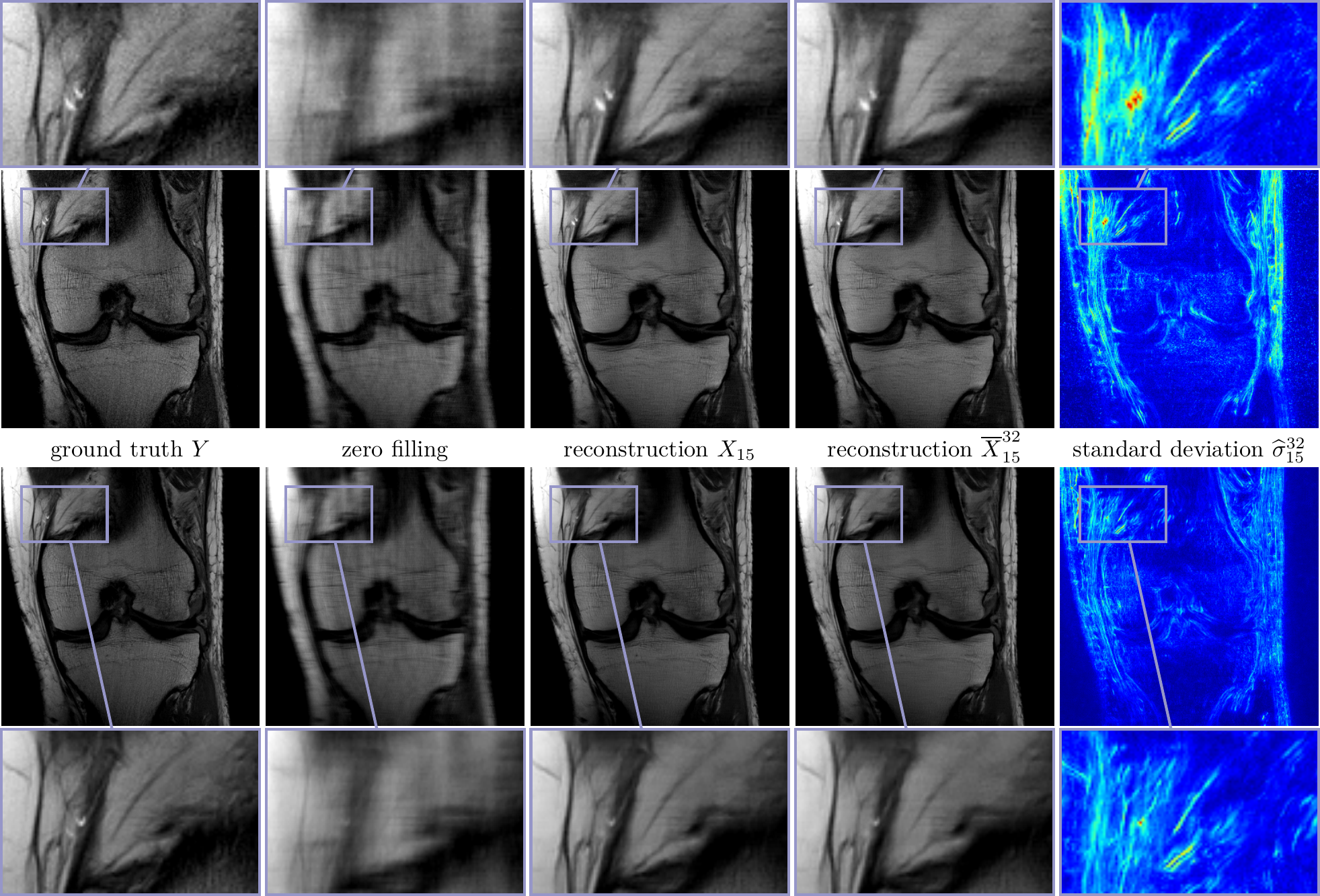}
\caption{Single (first row) and multi-coil (second row) MRI reconstruction results for PD data and $R=4$.
From left to right: ground truth images~$Y$, zero filling, deterministic reconstructions~$X_{15}$, stochastic reconstructions $\overline{X}_{15}^{32}$ and standard deviations~$\widehat{\sigma}_{15}^{32}$ ($0$ \protect\includegraphics[width=1.5cm,height=.2cm]{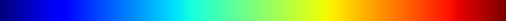} $0.02$).
}
\label{fig:PD4}

\end{figure*}

\Cref{fig:PD4} depicts two prototypic ground truth images of the PD data in the single (first row) and multi-coil (second row) case, the corresponding zero filling results, the deterministic~$X_{15}$ and the mean~$\overline{X}_{15}^{32}$ with $\beta=10^{-4}$ for $Q=1$ and $\beta=7.5\cdot 10^{-5}$ for $Q=15$ (see~\eqref{eq:meanDefintion}), and the standard deviation~$\widehat{\sigma}_{15}^{32}$ for an undersampling factor of~$R=4$.
The associated entropy levels are $\overline{H}(\Sigma)=-11.90$ and $\overline{H}(\Sigma)=-13.68$, respectively.
In both the deterministic and the stochastic reconstructions even fine details and structures are clearly visible, and the noise level is substantially reduced compared to the ground truth, which can be seen in the zoom with magnification factor~$3$.
Note that hardly any visual difference is observed in both reconstructions.
Clearly, the quality in the single-coil case is inferior to the multi-coil case.
Moreover, large values of the standard deviation are concentrated in regions with clearly pronounced texture patterns, which are caused by the lack of data in high-frequency $k$-space regions.
Thus, the standard deviation can be interpreted as a local measure for the epistemic uncertainty.
Since the proximal operator is applied after the update of the regularizer, high values of the standard deviation can only be found in regions where data is unknown. 

\begin{figure*}[htb]
\includegraphics[width=\linewidth]{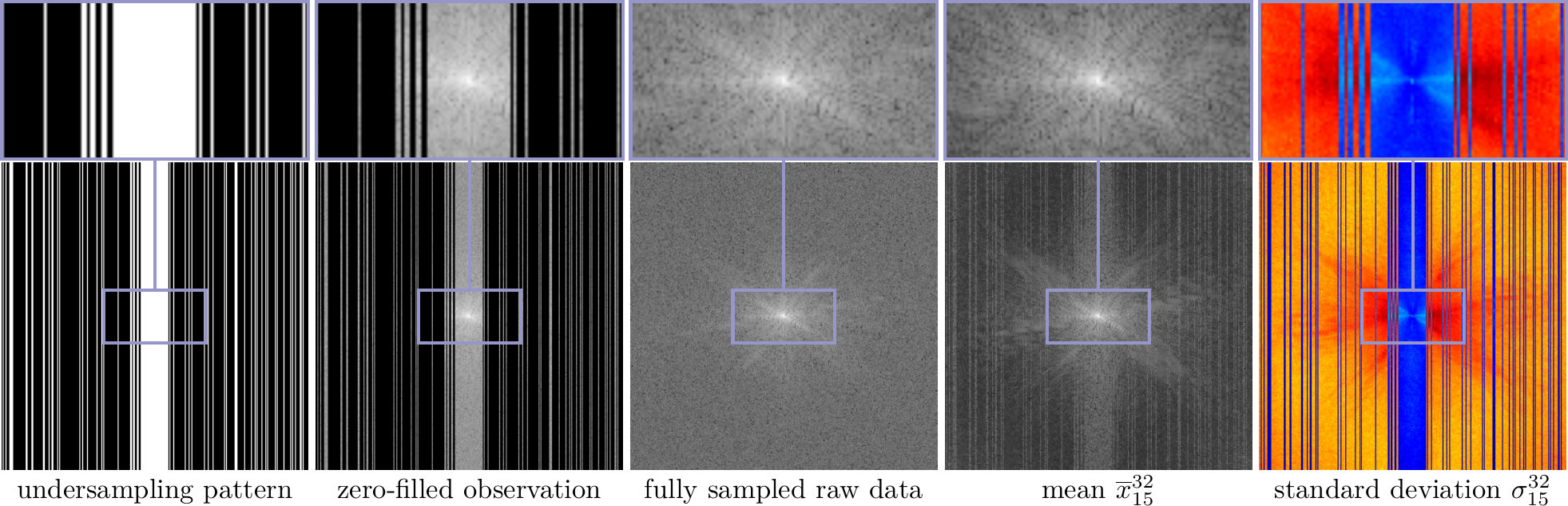}
\caption{
Visualization of the magnitude images in $k$-space (logarithmic scale) for $R=4$ in the single-coil case.
From left to right: undersampling pattern, zero-filled observation, fully sampled raw data, mean~$\overline{x}_{15}^{32}$, standard deviation~$\sigma_{15}^{32}$ ($-14.1$ \protect\includegraphics[width=1.5cm,height=.2cm]{jet.png} $-4.74$).}
\label{fig:kSpaceVisualization}
\end{figure*}

The $k$-space associated with the aforementioned single-coil case is depicted in \Cref{fig:kSpaceVisualization}.
In detail, the leftmost image visualizes the undersampling pattern resulting from the predefined Cartesian downsampling operator~$M_R$, which yields the zero-filled observation (second image) when combined with the fully sampled raw data (third image), where we plot the magnitudes in a logarithmic scale.
The fourth and the fifth image depict the mean~$\overline{x}_{15}^{32}$ and the standard deviation~$\sigma_{15}^{32}$ of the reconstruction in $k$-space.
As a result, our proposed method accurately retrieves the central star-shaped structures of the $k$-space representing essential image features, although the undersampling pattern is still clearly visible.
Moreover, the standard deviation peaks in the central star-shaped section when data is missing and thus empirically identifies regions with larger uncertainty.

\begin{figure*}[htb]
\includegraphics[width=\linewidth]{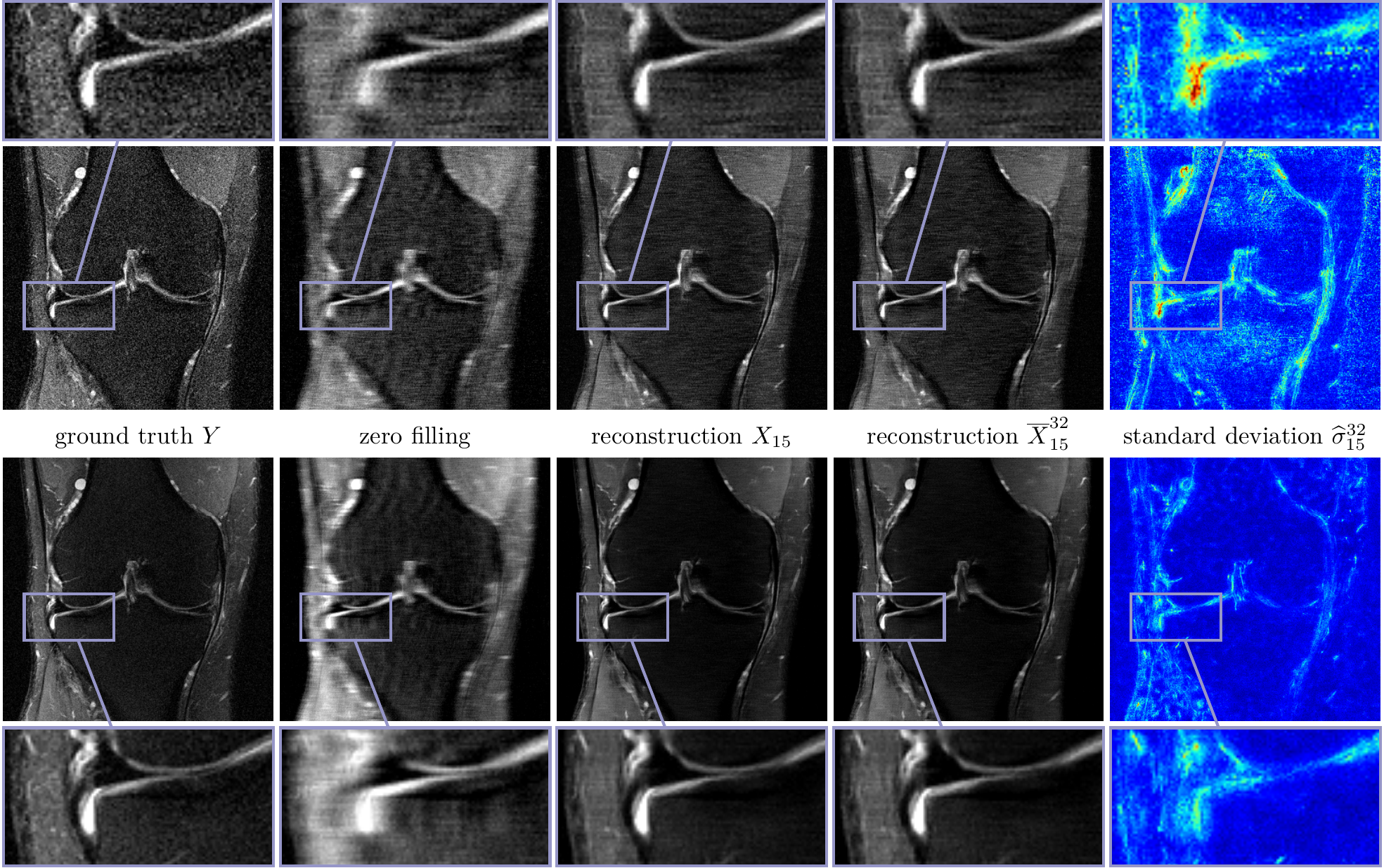}
\caption{Single (first row) and multi-coil (second row) MRI reconstruction results for PD-FS data and $R=4$.
From left to right: ground truth images~$Y$, zero filling, deterministic reconstructions~$X_{15}$, stochastic reconstructions $\overline{X}_{15}^{32}$ andstandard deviation~$\widehat{\sigma}_{15}^{32}$ ($0$ \protect\includegraphics[width=1.5cm,height=.2cm]{jet.png} $0.035$).
}
\label{fig:PD_FS4}
\end{figure*}

Likewise, \Cref{fig:PD_FS4} depicts the corresponding results for PD-FS data and $R=4$ using the same entropy levels as before, all other parameters are the same as in the previous \Cref{fig:PD4}.
We remark that the signal-to-noise ratio is smaller in PD-FS data than in PD data and thus the reconstructions have a tendency to include more noise and imperfections.
The inferior quality compared to PD is also reflected in the higher average intensities of the standard deviations.

\begin{figure*}[htb]
\includegraphics[width=\linewidth]{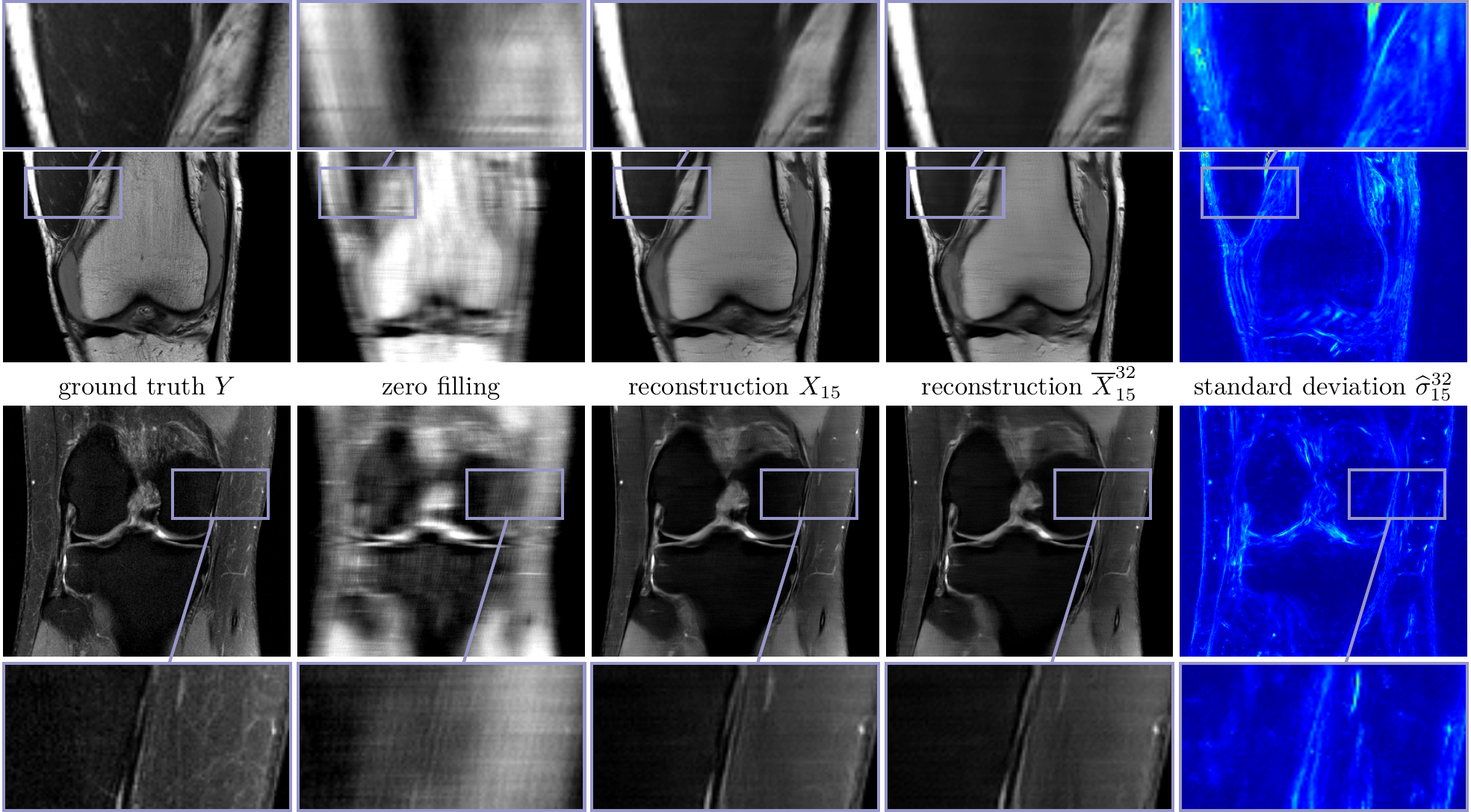}
\caption{
Multi-coil MRI reconstruction results for PD (first row) and PD-FS (second row) data and $R=8$.
From left to right: ground truth images~$Y$, zero filling, deterministic reconstructions~$X_{15}$, stochastic reconstructions $\overline{X}_{15}^{32}$ and standard deviation~$\widehat{\sigma}_{15}^{32}$ ($0$ \protect\includegraphics[width=1.5cm,height=.2cm]{jet.png} $0.02$).
}
\label{fig:multi8}
\end{figure*}

\begin{figure*}[htb]
\includegraphics[width=\linewidth]{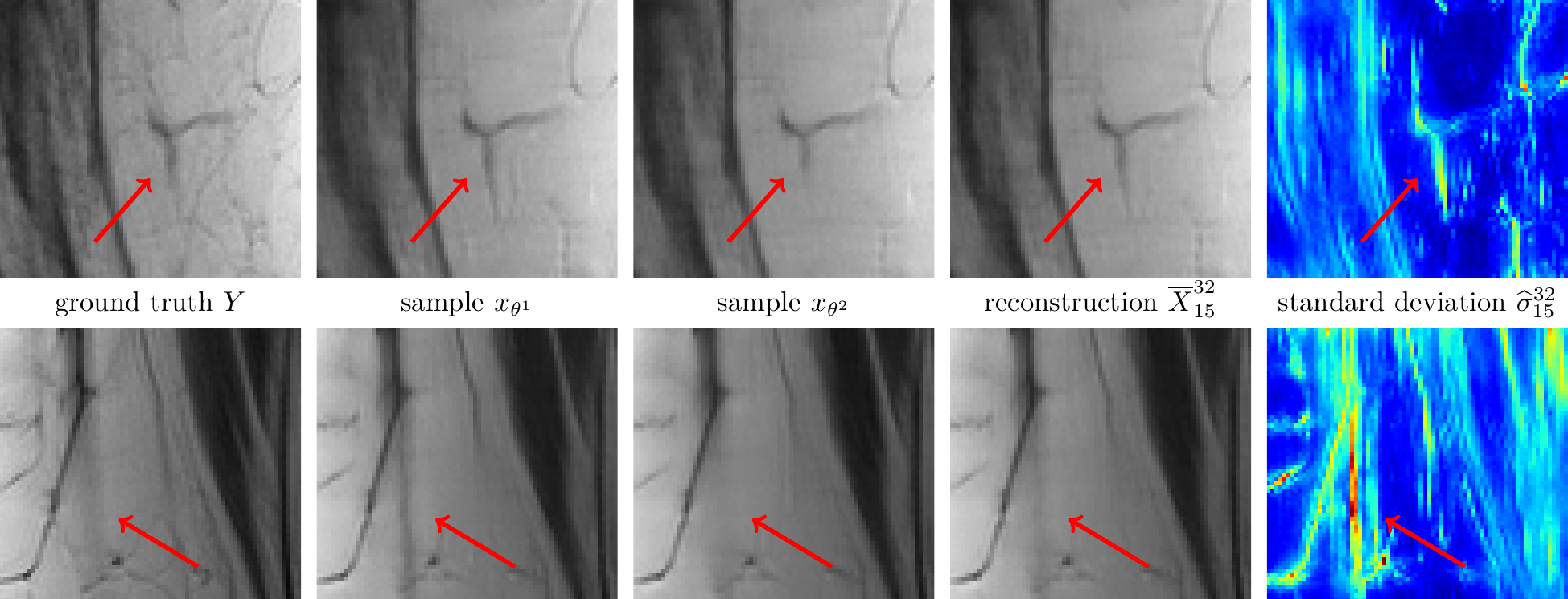}
\caption{
Zooms of multi-coil MRI reconstruction ($R=4$, PD).
From left to right: ground truth, two distinct samples, stochastic reconstruction and standard deviation
($0$ \protect\includegraphics[width=1.5cm,height=.2cm]{jet.png} $0.03$).
The arrows highlight patterns that are only visible in distinct samples.
}
\label{fig:singleRealizations}
\end{figure*}

\begin{figure*}[htb]
\includegraphics[width=\linewidth]{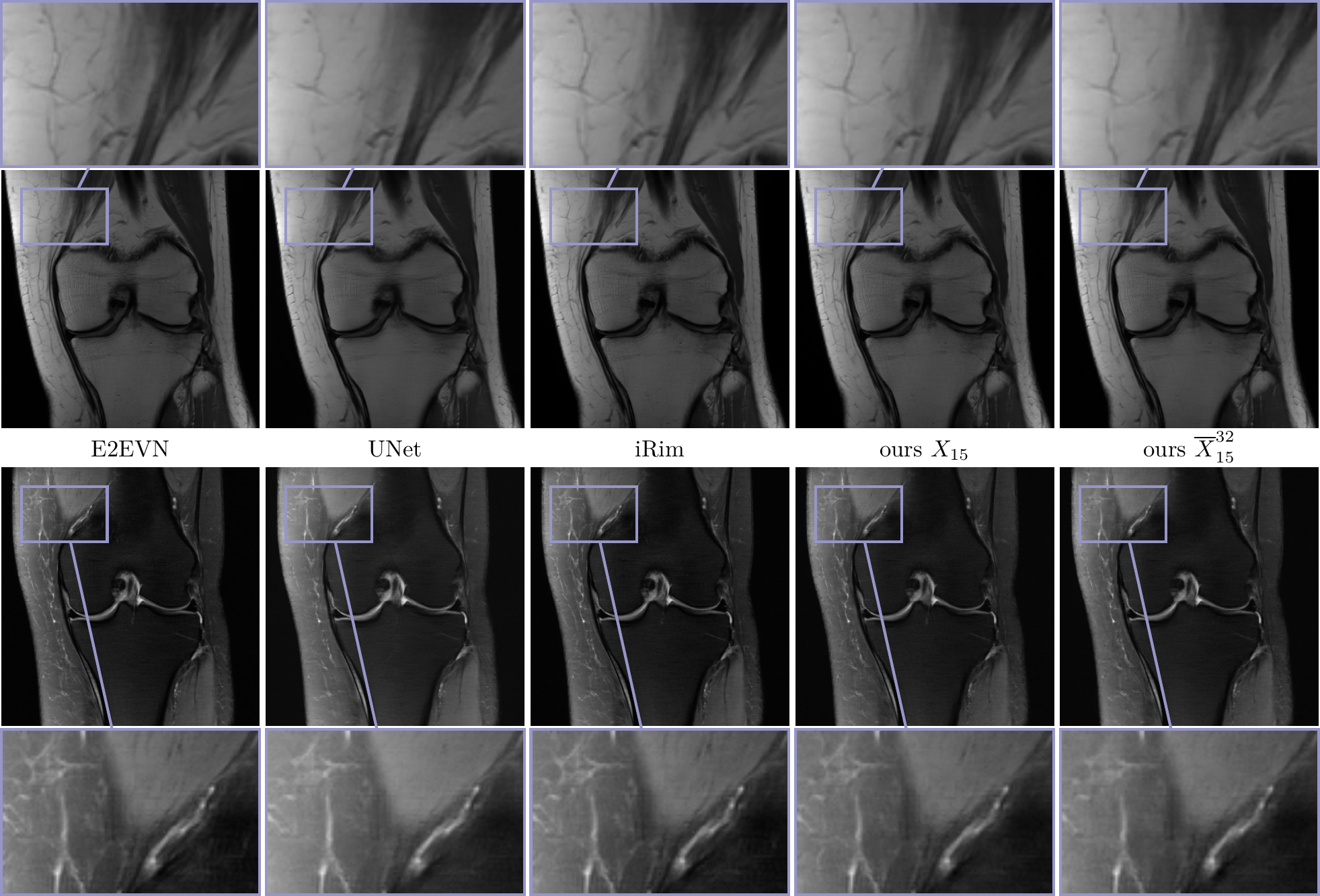}
\caption{
Visual comparison of selected MRI reconstruction methods for 
PD (first row) and PD-FS (second row), both with~$R=4$.
From left to right: E2EVN, UNet,  iRim, TDV(deterministic), TDV(stochastic).
}
\label{fig:viscomp}

\end{figure*}

\Cref{fig:multi8} shows the multi-coil reconstruction results for $8$-fold undersampling and both data sets in the same arrangement as before, the entropy level is $\overline{H}(\Sigma)=-31.41$.
As expected, the overall reconstruction quality is quantitatively and qualitatively inferior to the case $R=4$.
As before, the difference of the deterministic and the stochastic restored images is relatively small and the standard deviations properly identify regions with higher uncertainties.
Finally, \Cref{fig:singleRealizations} depicts zooms of two different MRI reconstructions ($R=4$, PD), in each row the ground truth, two realizations, the stochastic reconstruction and the standard deviation are visualized.
The regions highlighted by the arrows indicate structures and patterns that differ among various samples.
The variability of the single realizations can be interpreted as hallucinations, which are properly detected in the corresponding standard deviations.
This empirically validates that our proposed method to measure the standard deviation actually quantifies the magnitude of the model-related uncertainty.
\Cref{fig:viscomp} contains a visual comparison of our method with selected competitive methods from the fastMRI leader board.
As a result, both E2EVN~\cite{SrZb20} and iRim~\cite{PuKa20} achieve slightly superior quantitative results at the expense of significantly more learnable parameters.
In a qualitative comparison, we observe that our proposed method is capable of retrieving fine details, only the signal of a few high-frequency patterns is lost.
Finally, U-Net~\cite{KnZb20} results are inferior to the considered competitive methods -- both quantitatively and qualitatively.

\subsection{Covariance Matrices}
\begin{figure*}
\includegraphics[width=\linewidth]{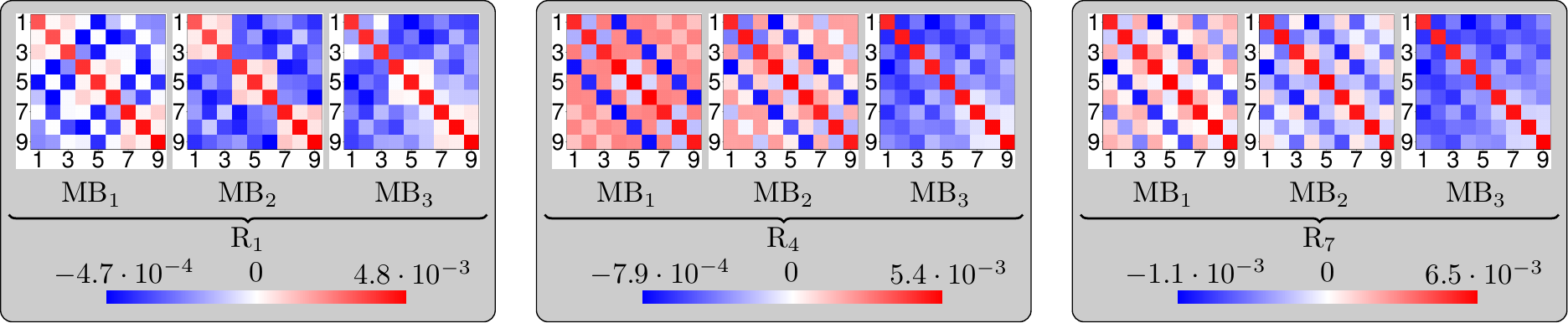}
\caption{
From left to right: triplets of color-coded covariance matrices of the convolution layers~$K_2$ in different macroblocks ($\textrm{MB}_i$ for $i=1,2,3$) and residual blocks ($\textrm{R}_j$ for $j=1,4,7$).
Note that we use different scalings for positive and negative values among each residual block.
}
\label{fig:COV}
\end{figure*}

\Cref{fig:COV} contains triplets of color-coded covariance matrices of the convolution layers~$K_2$ in different macroblocks and residual blocks (using the abbreviations $\textrm{MB}_i$ for $i=1,2,3$ and $\textrm{R}_j$ for $j=1,4,7$, respectively) in the multi-coil case with $\beta=7.5 \cdot 10^{-5}$.
Note that we use different scalings for positive and negative values among each residual block.
Each visualized covariance matrix is the mean of the individual covariance matrices of each convolution block~$K_2$ appearing in~$\Sigma$.
The resulting covariance matrices are clearly diagonally dominant with a similar magnitude of the diagonal entries among each residual block, but different magnitudes among different residual blocks.
Furthermore, most of the off-diagonal entries significantly differ from~$0$.
As a result, the entries of the covariance matrices associated with the first residual block in each macroblock have a tendency to smaller values compared to the ones of the last residual block.
Thus, the uncertainty of the network is primarily aggregated at latter residual blocks within each macroblock, which is in correspondence with error propagation theory: perturbations occurring shortly after the initial period have commonly a larger impact on a dynamical system than perturbations occurring later. 

\subsection{Eigenfunction Analysis}
Next, we perform a \emph{nonlinear eigenfunction analysis}~\cite{Gi18} following the approach in~\cite{EfKo19,KoEf20a} to heuristically identify local structures that are favorable in terms of energy.  
Classically, each pair $(v,\lambda)\in\C^n\backslash\{0\}\times\C$ of eigenfunction/eigenvalue for a given matrix~$A\in\R^{n,n}$ solves $Av=\lambda v$,
where the eigenvalue can be computed using the Rayleigh quotient $\frac{v^\ast Av}{v^\ast v}$.
Nonlinear eigenfunctions~$v$ for the matrix~$\nabla_v\mathcal{R}(v,\theta)$ satisfy $\nabla_v\mathcal{R}(v,\theta)=\Lambda(v)v$,
where the generalized Rayleigh quotient defining the corresponding \emph{eigenvalues} is given by
\[
\Lambda(v)=\frac{\langle\nabla_v\mathcal{R}(v,\theta),v\rangle}{\Vert x\Vert_2^2}.
\]
Thus, nonlinear eigenfunctions~$v$ are minimizers of the variational problem
\begin{equation}
\min_{v\in\C^{nQ}}\frac{1}{2}\Vert\nabla_v\mathcal{R}(v,\theta)-\Lambda(v)v\Vert_2^2
\label{eq:eigenfunctionAnalysis}
\end{equation}
subject to a fixed initial image.
We exploit Nesterov's projected gradient descent~\cite{Ne83} for the optimization in~\eqref{eq:eigenfunctionAnalysis}.
The nonlinear eigenfunctions locally reflect energetically minimal configurations and thus heuristically identify stable patterns that are favored by the model.

\begin{figure*}[htb]
\includegraphics[width=\linewidth]{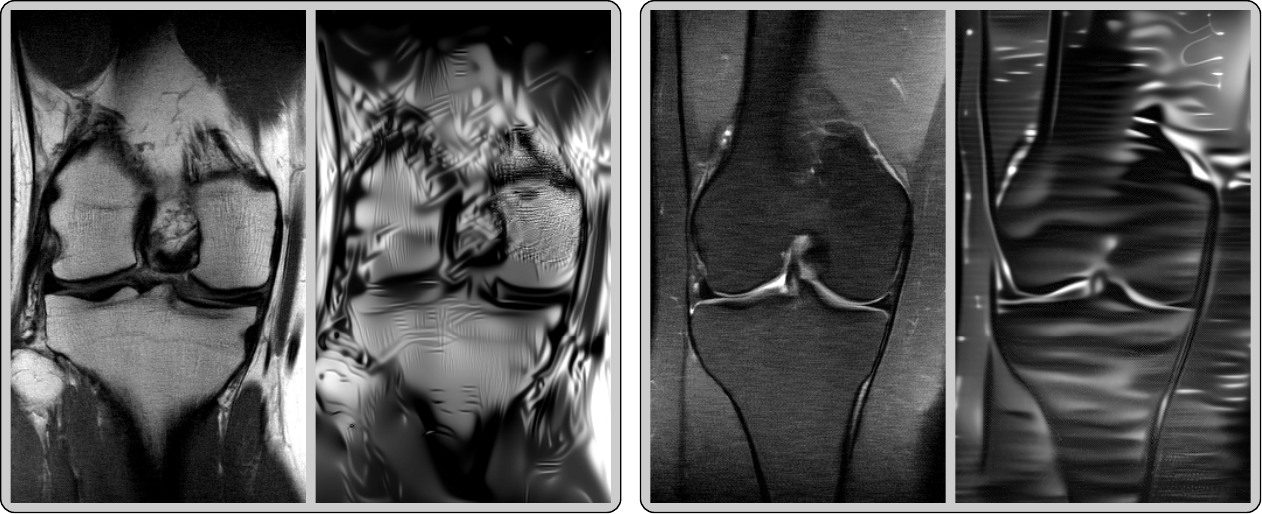}
\caption{Pairs of fully sampled initial images along with the corresponding eigenfunctions using PD (first pair) and PD-FS (second pair) as initialization.}
\label{fig:eigenfunctions}
\end{figure*}

\Cref{fig:eigenfunctions} depicts two pairs of reconstructed images~$X$ for the initialization and the corresponding root-sum-of-squares of the eigenfunctions in the deterministic single-coil case.
The resulting eigenfunctions predominantly exhibit piecewise smooth regions, where additional high-frequency stripe patterns and lines in the proximity of bone structures as well as blood vessels are hallucinated.
This behavior originates from two opposing effects: some backfolding artifacts caused by missing high-frequency components in $k$-space are removed in our approach, whereas certain high-frequency information are hallucinated.

\subsection{Effects of Entropy Level and Averaging}
\begin{figure*}[htb]
\centering
\includegraphics[width=.9\linewidth]{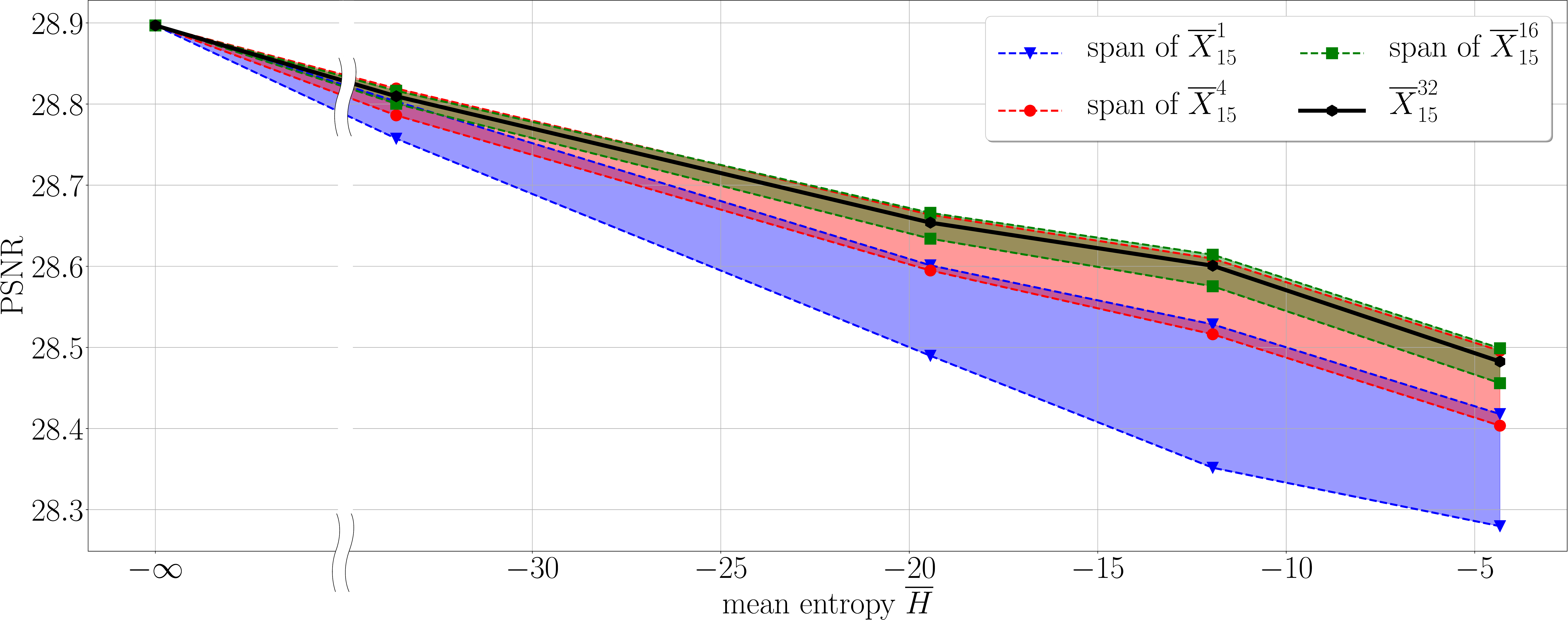}
\caption{Dependency of the PSNR value on the entropy and the averaging.}
\label{fig:ablationFull}
\end{figure*}

In the final experiment, we analyze the effects of the entropy level and the averaging on the PSNR values.
To this end, we draw $32$~instances $\underline{\theta}_{32}=(\theta_1,\ldots,\theta_{32})\in\Theta^{32}$ from $\mathcal{N}(\mu,\Sigma)$.
Then, for different levels of the entropy enforced by different values of~$\beta$ we calculate the lower and upper bounds of the PSNR values of $\overline{X}_{s}^{N}(z,T,\underline{\theta}_{N})$ for $1\leq {N}\leq 32$,
where $\underline{\theta}_{N}$ is any subset of $\underline{\theta}_{32}$ with ${N}$~elements.
\Cref{fig:ablationFull} depicts the resulting color-coded spans for ${N}~\in~\{1,4,16,32\}$ and five different levels of entropy (including the limiting case $\overline{H}(\Sigma)=-\infty$ in the deterministic case).
As a result, the PSNR curves monotonically decrease with higher levels of entropy, but even at the highest entropy level induced by $\beta=5\cdot 10^{-4}$ we observe only a relatively small decrease in the PSNR value.
Moreover, the spans of the different averaging processes clearly prove that higher values of~${N}$ are beneficial, that is why an averaging among a larger number of realizations should be conducted whenever possible.
Finally, we observe that the performance saturates with larger values of~${N}$.

\subsection{Limitations}
In the following we discuss potential limitations of our approach.
Since in each iteration the network parameters have to be drawn from the learned distribution, the training takes longer in the stochastic compared to the deterministic case.
Furthermore, to accurately quantify the uncertainty, N reconstructions have to be computed.
This leads to an N-fold increased reconstruction time compared to the deterministic scheme.
Higher levels of uncertainty result in a decrease of the PSNR score, as shown in \Cref{fig:ablationFull}. 
However, we believe that the advantage of having an estimate about the uncertainty outweighs the addressed limitations.

\section{Conclusion}
In this paper, we proposed a Bayesian framework for uncertainty quantification in single and multi-coil undersampled MRI reconstruction exploiting the total deep variation regularizer. 
To estimate the epistemic uncertainty, we introduced a stochastic optimal control problem, in which the weights of the regularizer are sampled from a learned multivariate Gaussian distribution.
With the proposed Bayesian framework, we can generate visually appealing reconstruction results alongside a pixelwise estimation of the epistemic uncertainty, which might aid medical scientists and clinicians to revise diagnoses based on structures clearly visible in the standard deviation plots.

\section{Appendix}
\subsection{Proximal map of the data fidelity}
\label{sec:DataProx}
To derive a closed-form expression of~$\prox_\mathcal{D}$, we first recall that the proximal map of $\mathcal{D}$ reads as
\[
\prox_{\tfrac{T}{S} \mathcal{D}}(\overline{x})=\argmin_{x}\underbrace{\frac{1}{2}\Vert x-\overline{x}\Vert_2^2+\frac{T}{2S}\Vert Ax-z\Vert_2^2}_{\eqqcolon G(x)}.
\]
We note that
\[
\nabla G(x)=x-\overline{x} + \tfrac{T}{S}A^\ast Ax-A^\ast z,
\]
which implies that the first-order optimality condition for~$G$ is
\[
\prox_{\tfrac{T}{S}\mathcal{D}}(\overline{x}) = (\Id + \tfrac{T}{S}A^*A)^{-1}(\overline{x}+A^*z).
\]
Taking into account $A=M_R F$ we can expand
\[
\Id+\tfrac{T}{S}A^\ast A=F^{-1}(\Id+\tfrac{T}{S}M_R^\ast M_R)F,
\]
where we exploited $F^\ast=F^{-1}$ since $F$ is unitary.
Thus, 
\[
\prox_{\tfrac{T}{S}\mathcal{D}}(\overline{x})
=F^{-1}(\Id+\tfrac{T}{S}M_R^\ast M_R)^{-1}F\Big(\overline{x}+\tfrac{T}{S}(F^{-1}M_R^\ast z)\Big),
\]
which ultimately leads to
\[
\prox_{\tfrac{T}{S}\mathcal{D}}(\overline{x})=F^{-1}((\Id+\tfrac{T}{S}M_R^\ast M_R)^{-1}(F\overline{x}+\tfrac{T}{S}M_R^\ast z)).
\]
Note that the inverse of the diagonal matrix can be computed very efficiently.

\subsection{Proximal map of Kullback--Leibler divergence}
\label{sec:KLDProx}
Next, we present a more detailed derivation of the proximal map~\cref{eq:proxDefinition} for 
\[
f(L)\coloneqq\beta(\alpha\tr(LL^\top)-\log(\det(LL^\top)))
\]
appearing in \eqref{eq:multivariateKLD}.
We first observe that 
\begin{align*}
\tr(LL^{\top})&=\sum_{i,j} L_{ij}L_{ij}=\sum_{i,j} L_{ij}^2,\\
\det(LL^{\top})&=\det(L)\det(L^{\top})=\det(L)^2=\prod_{i}L_{ii}^2.
\end{align*}
Recall that $L$ admits a block diagonal structure, in which each block $l_1,\ldots,l_U\in\R^{o^2\times o^2}$ is a regular lower triangular matrix.
A straightforward computation reveals $f(L)=\sum_{u=1}^U f(l_u)$.
Thus, we can restrict to a single block matrix~$l$ and rewrite $f$ as follows:
\[
f(l)=\alpha\beta\sum_{i,j=1}^p l_{ij}^2-2\beta\sum_{i=1}^p\log(l_{ii}).
\]
The proximal map of the function $f$ reads as
\[
\prox_{h f}(\overline{l})=\argmin_{l}\left\{E(l)\coloneqq\tfrac{1}{2 h}\Vert l-\overline{l}\Vert_2^2+f(l)\right\},
\]
where the minimum is taken among all lower triangular and regular matrices.
The gradient of $E$ is given by
\[
(\nabla E(l))_{ab}=
\begin{cases}
\displaystyle l_{aa}^2(1+2\alpha\beta h)-\overline{l}_{aa}l_{aa}-2\alpha h,& a=b,\\[1em]
2\alpha\beta h l_{ab}+l_{ab}-\overline{l}_{ab},& a\neq b.
\end{cases}
\]
Thus, the optimization problem~\eqref{eq:proxDefinition} can be optimized component-wise and results in a quadratic equation.
Overall, a closed-form expression for~$f$ reads as
\[
\prox_{ h f}(\overline{l})_{ab}=
\begin{cases}
\displaystyle\frac{\overline{l}_{aa}+\sqrt{\overline{l}_{aa}^2+8\beta h(1+2\alpha\beta h)}}{2(1+2\alpha\beta h)}
,& a=b,\\[1em]
(1+2\alpha\beta h)^{-1}\overline{l}_{ab},& a\neq b.
\end{cases}
\]
The expression for the proximal map of $f$ for the entire matrix~$L$ is given as the concatenation of the proximal maps for the individual blocks.

\FloatBarrier

\end{document}